\documentclass[a4paper,11pt]{article}
\pdfoutput=1 
\usepackage{jheppub}
\usepackage[T1]{fontenc}
\usepackage{color}
\usepackage{slashed}
\usepackage{soul}
\usepackage{amsmath}
\newcommand{\bef}{\begin{figure}[hbt]\centering}
\newcommand{\eef}{\end{figure}}
\allowdisplaybreaks

\usepackage{graphicx}

\usepackage[numbers, sort&compress]{natbib}

\title{\boldmath One-Point energy correlator inside jets}

\author{Zihao Mi,}
\author{Zhan Wang}

\affiliation{Department of Physics, Beijing Normal University, Beijing 100875, China }

\emailAdd{zihaomi@mail.bnu.edu.cn}
\emailAdd{zhanwang@mail.bnu.edu.cn}

\abstract{In this work, we introduce a new jet observable, the one-point energy correlators (EC), designed to characterize the in-jet energy flow distribution by measuring energy deposition at a specific angle relative to the jet axis. Building upon the transverse momentum dependent physics, we aim for the EC to provide novel insights into jet substructure and offer a new approach to study TMD physics, particularly gluon transverse momentum dependent fragmentation functions (TMDFFs) which are notoriously difficult to extract. We obtain the factorization of the EC jet function within Soft-Collinear Effective Theory and leverage the framework of semi-inclusive TMD fragmenting jet functions. We resum large global logarithms and non-global logarithms (NGLs) and show that the normalized EC jet function exhibits significantly reduced dependence on the factorization scale and is primarily sensitive to the jet scale. Finally, after incorporating non-perturbative effects, we present numerical calculations up to NNLL accuracy for global logarithms and LL accuracy for NGLs, and we compare these predictions with PYTHIA 8 simulations.}

\begin{document}
\maketitle
\flushbottom

\section{Introduction}
\label{sec:intro}
The Large Hadron Collider (LHC)~\cite{ATLAS:2008xda} has ushered in an era of unprecedented experimental precision in high-energy physics. Experiments like ATLAS and CMS have collected vast datasets, enabling detailed measurements of various final states, including the production of jets, photons, massive vector bosons, and heavy quarks, across a wide range of kinematic variables. The quarks and gluons produced in hard interactions are not observed directly due to color confinement, instead, they fragment and hadronize into observable collimated sprays of particles known as jets. Jets act as experimental proxies for the initial partons, and their internal structure that the pattern of energy and particle flow within them, offers a rich tapestry of information about the fundamental dynamics of Quantum Chromodynamics (QCD), including parton showering and the transition to hadrons. The field of jet substructure employs a suite of observables, including jet mass~\cite{Chien:2012ur,ATLAS:2012am}, shapes~~\cite{Chien:2014nsa,Hornig:2016ahz,Seymour:1997kj,Ellis:2010rwa}, N-jettiness~~\cite{Stewart:2010tn,Lesser:2023nsy}, and energy correlators~~\cite{Basham:1978bw,Basham:1978zq,Jaarsma:2023ell,Hofman:2008ar,Belitsky:2013ofa,Belitsky:2013xxa,Kologlu:2019mfz,Korchemsky:2019nzm,Dixon:2019uzg,Chen:2020vvp,moult2025energycorrelatorsjourneytheory}, to analyze these patterns, aiming to disentangle perturbative and non-perturbative contributions and probe the underlying physics with increasing theoretical rigor.

Among the suite of jet substructure observables, energy-energy correlators (EEC) have recently gained prominence. First introduced decades ago in the context of $e^{+}e^{-}$ annihilation~\cite{Basham:1978bw}, EEC quantify the angular correlations of energy deposition within a collision event. They are typically defined as distributions weighted by the product of the energies of pairs of particles as a function of the angular distance between them. They are relatively robust against contamination from soft background radiation. Furthermore, EEC exhibit a remarkable theoretical tractability, connecting directly to correlation functions of fundamental field theory operators~~\cite{Sveshnikov:1995vi,Tkachov:1995kk,Korchemsky:1999kt,Bauer:2008dt,Kravchuk:2018htv,Belitsky:2013bja}, this operators also known as ANEC operators. This connection facilitates the application of advanced theoretical techniques and provides a clear interpretation of the measurements in terms of QCD dynamics.

A significant and rapidly developing area of research lies at the intersection of jet substructure~\cite{Seymour:1997kj,Lee:2022kdn,Ellis:1992qq,Ellis:2010rwa,Jain:2011xz,Jain:2011iu,Kaufmann:2015hma,Kaufmann:2016nux,Kolodrubetz:2016dzb,Kang:2017glf,Kang:2016mcy,Kang:2016ehg,Kang:2017mda} and transverse momentum dependent (TMD) physics~\cite{Collins:1981uw,Collins:1981va}. Specifically, research explores whether the detailed information encoded within the angular patterns of energy flow inside jets, as probed by EEC, can be utilized to infer or constrain the properties of TMDPDFs and FFs, such as the Sivers function~\cite{Kang:2011hk} and transversity, provide a multi-dimensional picture of the quark and gluon distribution inside hadrons, incorporating both their transverse momentum and spin. These functions are crucial for understanding the nucleon's spin puzzle and probing the complex interplay between partonic orbital angular momentum and spin~\cite{Avakian:2008dz}. This connection represents an active and challenging frontier in contemporary particle and nuclear physics, with previous work~\cite{Kang:2017btw,Yuan:2007nd} exploring similar ideas of hadron production inside jets by examining Collins azimuthal asymmetries.

\begin{figure}[htbp]
	\centering
	\includegraphics[width=0.5\textwidth]{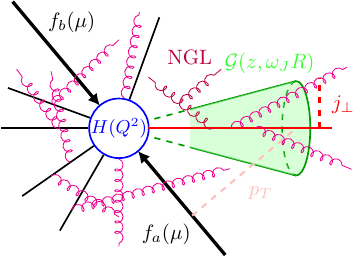}
	\caption{Schematic of the one-point energy correlator inside jets: it measures energy flow inside a jet at a fixed angle $\chi$ from the jet axis and its TMD factorization.}
	\label{fig:1}
\end{figure}

In this work, building upon the framework of TMDs, we introduce a new observable, the one-point energy correlators (EC). It characterizes the in-jet energy flow distribution by measuring the energy deposited at a specific angle relative to the jet's axis as Fig~\ref{fig:1}.
The motivation for integrating these fields is compelling. It promises novel perspectives on fundamental QCD dynamics, parton fragmentation processes, and the intricate three-dimensions structure of hadrons. This confluence is driven by complementary needs and capabilities from both fields. TMD physics requires new, theoretically clean observables to reliably extract the non-perturbative TMD functions, which are notoriously difficult to isolate, the inclusive hadron summation combined with momentum sum rules eliminates the dependence on collinear fragmentation functions~\cite{Moult:2018jzp}. Simultaneously, the theoretical framework for EEC, particularly when analyzed in specific kinematic limits like the back-to-back or near-collinear regimes, naturally connects to the factorization theorems underpinning TMD physics~\cite{Kang:2017glf}. EEC thus emerge as a potentially powerful new tool to address the long-standing challenge of precisely determining TMDs and testing their fundamental properties, such as evolution and universality. This offers a novel approach, via EC, to study TMD physics independently of fragmentation effects, especially for gluon TMDFFs.

The remainder of this paper is organized as follows. In Section~\ref{sec:fac}, we review EEC and give the definition of the EC, then derive the factorization for the EC. Section~\ref{sec:resum} presents the resummation of large and non-global logarithms and discusses non-perturbative corrections. Numerical results at $\text{NNLL}\otimes\text{NP}+\text{LL}_{\text{NG}}$ accuracy and comparisons with PYTHIA~8 are given in Section~\ref{sec:results}. We conclude in Section~\ref{sec:conclude}.

\section{The Factorization Theorem}
\label{sec:fac}
In this section, we first review energy correlators observables and the TMD formalism. We then synthesize these frameworks to derive a factorization theorem for the energy correlators of hadrons measured within a jet, specifically targeting the collinear limit.

\subsection{Energy Correlators}
\label{subsec:eec}
Energy correlators are a class of observables that probe correlations in the asymptotic energy  flux in collider experiments. The two-point energy correlator was first defined in $e^{+}e^{-}$ collisions  as~\cite{Basham:1978zq}
\begin{align}
	\text{EEC}(\chi)=\frac{1}{\sigma} \sum_{a,b} \int \frac{E_a E_b}{Q^2} d\sigma_{e^-e^+\rightarrow h_a h_b X} \, \delta(\chi - \theta_{ab}) \,,
\end{align}
Here, the sum runs over all the hadron pairs and the cross section is weighted by the energy fraction, which is the product of the energies of hadron normalized by the center-of-mass energy of the system. The $\theta_{ab}$ is the angle between the two hadrons.

This fundamental definition can be straightforwardly generalized to multi-point energy correlation functions or one-point energy correlation. From a field-theoretic perspective, energy correlators offer a significant advantage as they can be directly formulated in terms of matrix elements of well-defined energy flow operators~\cite{Hofman:2008ar}
\begin{align}
	\mathcal{E}(\hat{n}) = \lim_{r \to \infty} r^2 \int_0^{\infty} \, dt \, \, n^i \, T_{0i}(t, \hat{n}) \,.
\end{align}

Analogously, for analyzing the internal structure of jets, we define a one-point energy correlator for hadrons within jets as
\begin{align}
	\text{EC}(\chi)=\frac{1}{\sigma} \sum_{a} \int \frac{\omega_h}{\omega_J}\, d\sigma_{e^-e^+\rightarrow (\text{jeth}) X} \, \delta(\chi - \theta_{h}) \,,
	\label{eq:definition}
\end{align}
where the sum includes all hadrons $h$ observed within reconstructed jets, $\omega_h$ is the energy of the hadron, and $\omega_J$ is the energy of the jet that contains this hadron. The observable is therefore weighted by the hadron's energy fraction $z_h$ $(\omega_h / \omega_J )$ relative to the jet, and $\theta_{h}$ is the angle between the hadron and the standard jet axis~\cite{Neill:2016vbi}.

\subsection{Factorization Of The Energy Correlator Jet Function}
The transverse momentum of hadron within a jet, measured relative to the standard jet axis, is parametrically smaller than the characteristic jet scale. Consequently, a TMD factorization framework is necessary for perturbative calculation. The TMD region is defined by the hierarchy of scales $\Lambda_{\text{QCD}} \lesssim j_\perp \ll p_T R$ . Within this region, the angle $\chi$ between the hadron and the jet axis is small and can be approximated as $\chi \sim \frac{2j_{\perp}}{z_{h}\omega_{J}} $ . Following the definition of EC~\eqref{eq:definition}, the EC jet function can be constructed from the siTMDFJF, denoted as $\mathcal{G}_c^h$ . This is achieved by integrating the siTMDFJF over the hadron's transverse momentum, with the $z_h$ weighting and angular constraint
\begin{align}
	J_{c}(z_{c},\chi,\omega_{J}R,\mu_{J}) =\sum_{h}\int d^2\boldsymbol{j}_{\perp}\int dz_{h}z_{h}\mathcal{G}_{c}^{h}(z_{c},z_{h},\omega_{J}R,j_{\perp},\mu_{J})\delta\bigg(\frac{2j_{\perp}}{z_{h}\omega_{J}}-\chi\bigg)\,,
    \label{eq:fac}
\end{align}
where $\mu_J$ represents the characteristic scale of the jet function. The siTMDFJF itself can be factorized within Soft-Collinear Effective Theory (SCET)~\cite{Kang:2017glf,Bauer:2000ew,Bauer:2000yr,Bauer:2001ct,Bauer:2001yt,Bauer:2002nz}
\begin{align}
	\label{eq:GG}
	\mathcal{G}_c^h(z,z_h,\omega_JR,\boldsymbol{j}_\perp,\mu)=
	 & \mathcal{H}_{c\to i}(z,\omega_{J}R,\mu)\int d^{2}\boldsymbol{k}_{\perp}d^{2}\boldsymbol{\lambda}_{\perp}\delta^{2}\left(z_{h}\boldsymbol{\lambda}_{\perp}+\boldsymbol{k}_{\perp}-\boldsymbol{j}_{\perp}\right) \nonumber
	\\
	 & \times D_{h/i}(z_h,\boldsymbol{k}_\perp,\mu,\nu)S_i(\boldsymbol{\lambda}_\perp,\mu,\nu R)\,.
\end{align}
Here, the sum extends over all possible intermediate parton $i$ that fragment into the $h$ . The components of the factorization formula are
\begin{itemize}
	\item $\mathcal{H}_{c\to i}(z,\omega_{J}R,\mu)$ are hard functions. They describe the transition from an energetic parton $c$ (originating from the hard scattering process) to parton $i$, parton $i$ subsequently initiates a jet with a large light-cone momentum component $\omega_J$ and radius $R$, and carries a momentum fraction $z$ with respect to parton $c$. These functions effectively incorporate the effects of radiative emissions that occur outside the reconstructed jet~\cite{Kang:2016ehg}.
	\item $S_i(\boldsymbol{\lambda}_{\perp}, \mu, \nu R)$ are the in-jet soft functions, which account for soft radiation with transverse momentum $\lambda_\perp$ (relative to the jet axis) inside the jet of radius $R$. The detailed derivation will be found in~\cite{Kang:2017mda}.
	\item $D_{h/i}(z_h, \boldsymbol{k}_{\perp}, \mu, \nu)$ are the usual TMDFFs. They describe the fragmentation of parton $i$ into the observed hadron $h$ , which carries a longitudinal momentum fraction $z_h$ (relative to parton $i$) and a transverse momentum $\boldsymbol{k_\perp}$ . These functions characterize the non-perturbative hadronization process within the TMD regime and are primarily sensitive to scales much smaller than the jet radius, thus lacking $R$ dependence.
\end{itemize}
All constituent functions in this factorization formula have been calculated up to NLO and are rapidity regularized~\cite{Kang:2017glf}, the rapidity scale $\nu$ will cancel. This precision is crucial for determining their renormalization group (RG) evolution properties and natural scale. As will be detailed in Section~\ref{sec:resum}, knowledge of these RG equations enables the resummation of potentially large logarithms, such as $\ln R$ and $\ln(p_T R/ \chi)$. This resummation is systematically achieved by evolving each component from its natural scale to the common hard scale $\mu \sim p_T$, at which the overall observable is evaluated.

lots of jet substructure measurements~\cite{CMS:2011ab,ATLAS:2011juz,CMS:2012oyn,ATLAS:2012am,ALICE:2014dla,ATLAS:2011myc,CMS:2014jjt,ALICE:2013dpt} are usually performed for inclusive jet production, we now consider an inclusive jet cross-section for $p p \rightarrow \text{jet} X$ where a hadron $h$ is identified inside the reconstructed jet, and its contribution to the EC is measured. Following the framework detailed in~\cite{Kang:2016mcy,Ellis:1978sf,Ellis:1978ty,Collins:1981ta,Collins:1989gx}, the factorization theorem for the normalized EC distribution in the process $p p \rightarrow \text{jet} + X$ can be written as
\begin{align}
	\operatorname{EC}(\chi)
	 & =\frac{1}{\sigma_0}\sum_{a,b,c}\int  \frac{2}{p_T} d p_T\int d \eta \int_{z_{0}}^{1}\frac{1}{z_{c}^{2}} \int_{VW/z_c}^{1-(1-V)/z_c}dv \int_{VW/vz_c}^1 d\omega
	\nonumber                                                                                                                                                         \\
	 & \times x_a f_a(x_a,\mu) x_b f_b(x_b,\mu) \frac{d\sigma^c_{ab}}{dvdw} J_c(z_c,\chi,\omega_JR,\mu) \,.
\end{align}
The integration variables $p_T$ and $\eta$ are the transverse momentum and pseudo rapidity of the jet. The kinematic variables appearing in the integration limits are defined as $z_0 \equiv 1 -V+VW$ , and the hadron-level kinematic variables are defined as~\cite{Jager:2002xm}
\begin{align}
	V\equiv1+\frac{T}{S},\quad W\equiv\frac{-U}{S+T},\quad S\equiv(P_A+P_B)^2,\quad T\equiv(P_A-P_J)^2,\quad U\equiv(P_B-P_J)^2 \,,
\end{align}
the corresponding parton-level kinematic variables are
\begin{align}
	v\equiv1+\frac{t}{s},\quad w\equiv\frac{-u}{s+t},\quad s\equiv(p_a+p_b)^2,\quad t\equiv(p_a-p_c)^2,\quad u\equiv(p_b-p_c)^2 \,,
\end{align}
where $f_a$ and $f_b$ denote the parton distribution functions in the proton, carrying longitudinal momentum fractions $x_a$ and $x_b$ respectively, at factorization scale $\mu$. As discussed in~\cite{Kang:2017glf}, this particular choice of observable and kinematics offers an opportunity to disentangle the effects of TMDPDFs from those of TMDFFs, the transverse momentum spectrum within jets only depends on TMDFFs in $pp$ collision. Crucially, if the observable primarily reflects the transverse momentum distribution within the jet, and this distribution is solely governed by TMDFFs, the observable itself should exhibit no significant direct dependence on TMDPDFs. This feature would allow for a more direct extraction and constraint of TMDFFs.

\section{Resummation and Non-perturbative Corrections}
\label{sec:resum}
Given that the EC jet function  is defined inclusively over the hadron species via the siTMDFJF, their resummation structures are inherently similar. We will therefore develop the resummation formalism for the EC Jet function, building upon the established framework for siTMDFJFs~\cite{Kang:2017glf} and consider the theoretical precision of EC. Furthermore, this section addresses two key phenomena: non-global logarithms (NGLs), which are pertinent as the EC Jet function is a non-global observable, and non-perturbative hadronization effects.
\subsection{Resummation}
The resummation of large logarithms is performed by evolving each component of the factorized cross section from its natural scale, where such logarithms are minimized, to a common hard scale $\mu_H$. The natural scales for the TMDFFs, in-jet function, hard function and EC jet function are chosen as
\begin{align}
	\mu_D \sim \mu_b \,,\quad \nu_D \sim \omega_J \,, \quad \mu_S \sim \mu_b \,, \quad \nu_S \sim \frac{\mu_b}{\tan(R/2)} \,, \quad \mu_H \sim p_T \,, \quad \mu_J \sim p_T R \,,
\end{align}
the scale $\mu_b$ which is defined as $\mu_b = 2e^{-\gamma_E}/b$, and the resummation of large logarithms are performed in $b$-space. The RG and RRG evolution equations for the corresponding renormalization scale $\mu$ and rapidity scale $\nu$, respectively, can then be readily obtained. For the TMDFFs $D_{h/i}(z_h, \boldsymbol{b}, \mu, \nu)$ and the in-jet soft function ${S}(\boldsymbol{b}, \mu, \nu)$
\begin{subequations}
	\begin{align}
		\frac{d}{d\ln \mu} \ln D_{h/i}(z_h, \boldsymbol{b}, \mu, \nu) & = \gamma_{D}\bigg(\frac{\nu}{\omega_J}\text{;} \,\alpha_s{(\mu)}\bigg) \,,
		\\[.2cm]
		\frac{d}{d\ln \mu} \ln {S}(\boldsymbol{b}, \mu, \nu)          & = \gamma_{S}\bigg(\frac{\nu\tan(R/2)}{\mu}\text{;}\,\alpha_s{(\mu)}\bigg) \,,
		\\[.2cm]
		\frac{d}{d\ln \nu} \ln D_{h/i}(z_h, \boldsymbol{b}, \mu, \nu) & = -\gamma_R\bigg(\frac{\mu}{\mu_b}\text{;}\,\alpha_s{(\mu)}\bigg) \,,
		\\[.2cm]
		\frac{d}{d\ln \nu} \ln {S}(\boldsymbol{b}, \mu, \nu)          & = \gamma_R\bigg(\frac{\mu}{\mu_b}\text{;}\,\alpha_s{(\mu)}\bigg) \,.
	\end{align}
\end{subequations}
The evolution equation for the hard function $\mathcal{H}_{i\to j}(z,\omega_JR,\mu)$ is
\begin{align}
	\frac{d}{d\ln\mu}\mathcal{H}_{i\to j}(z,\omega_JR,\mu)=\sum_k\int_z^1\frac{dz^{\prime}}{z^{\prime}}\gamma_{ik}\left(\frac{z}{z^{\prime}},\frac{\omega_J\tan(R/2)}{\mu};\,\alpha_s{(\mu)}\right)\mathcal{H}_{k\to j}(z^{\prime},\omega_JR,\mu) \,.
\end{align}
The relevant anomalous dimensions are given by
\begin{subequations}
	\begin{align}
		\gamma_{D}\bigg(\frac{\nu}{\omega_J};\alpha_s{(\mu)}\bigg)             & =\Gamma_{\mathrm{cusp}}[\alpha_s{(\mu)}]\ln\left(\frac{\nu^2}{\omega_J^2}\right)+\gamma^{D}[\alpha_s{(\mu)}\big] \,,
		\\[.2cm]
		\gamma_{S}\bigg(\frac{\nu\tan(R/2)}{\mu};\alpha_s{(\mu)}\bigg)         & =-\Gamma_{\mathrm{cusp}}[\alpha_s{(\mu)}]\ln\left(\frac{\nu^2 \tan^2(R/2)}{\mu^2}\right)+\gamma^{S}[\alpha_s{(\mu)}\big] \,,
		\\[.2cm]
		\gamma_{ij}\bigg(z,\frac{\omega_J\tan(R/2)}{\mu};\alpha_s{(\mu)}\bigg) & =\Gamma_{\mathrm{cusp}}[\alpha_s{(\mu)}]\ln\left(\frac{\omega_J^2\tan^2(R/2)}{\mu^2}\right)-\gamma^{D}[\alpha_s{(\mu)}]
		\\[.2cm]
		                                                                       & -\gamma^{S}[\alpha_s{(\mu)}]+\frac{\alpha_s{(\mu)}}{\pi}P_{ji}(z) \,,
		\\[.2cm]
		\gamma_R\bigg(\frac{\mu}{\mu_b};\alpha_s{(\mu)}\bigg)                  & =-\int_{\mu_b}^{\mu}\frac{d\mu^\prime}{\mu^\prime}\Gamma_{\mathrm{cusp}}[\alpha_s{(\mu^\prime)}]+\gamma^R[\alpha_s{(\mu_b)}] \,.
	\end{align}
\end{subequations}
The coefficients for these anomalous dimensions are provided in Appendix~\ref{app:A}. A common evolution rapidity scale $\nu $ is chosen for the TMDFFs and the in-jet soft function. The renormalized TMDFFs and the in-jet soft function can then be expressed as
\begin{align}
	D_{h/i}(z_h,\boldsymbol{b},\mu,\nu) & =D_{h/i}(z_h,b,\mu_b,\nu_D)\left(\frac{\nu}{\nu_D}\right)^{-\gamma^R[\alpha_s(\mu_b)]}
	\nonumber                                                                                                                                                                                                                                                                 \\
	                                    & \times\exp\Bigg\{\int_{\mu_b}^{\mu}\frac{d\mu^\prime}{\mu^\prime}\bigg[\Gamma_{\mathrm{cusp}}[\alpha_s(\mu^{\prime})]\ln\left(\frac{\nu^2}{Q^2}\right)+\gamma^{D}[\alpha_s(\mu^{\prime})]\bigg]\Bigg\}\,,
	\\
	S_i(\boldsymbol{b},\mu,\nu R)=
	                                    & S_i(b,\mu_b,\nu_{S}R)\left(\frac{\nu}{\nu_{S}}\right)^{\gamma^R(\alpha_s(\mu_b))}
	\nonumber                                                                                                                                                                                                                                                                 \\
	                                    & \times\exp\Bigg\{\int_{\mu_b}^{\mu}\frac{d\mu^\prime}{\mu^\prime}\bigg[-\Gamma_{\mathrm{cusp}}[\alpha_s(\mu^{\prime})]\ln\left(\frac{\nu^2 \tan^2(R/2)}{{\mu^{\prime}}^2}\right)+\gamma^{S}[\alpha_s(\mu^{\prime})]\bigg]\Bigg\}\,.
\end{align}
The evolution of the hard function, $\mathcal{H}_{i \to j}(z, \omega_J R, \mu)$, involves separating its anomalous dimension into a purely diagonal part and the Altarelli-Parisi splitting functions. We can separate these two parts of the anomalous dimensions, then the hard function can then be expressed by factoring out this RG evolution from an initial scale $\mu_J$ to the scale $\mu$
\begin{align}
	\mathcal{H}_{i \to j} & (z, \omega_J R, \mu) =
	\nonumber                                                                                                                                                                                                                                       \\
	                      & \exp \left( \int_{\mu_J}^{\mu} \frac{d\mu'}{\mu'} \, \Gamma_{\mathrm{cusp}}[\alpha_s{(\mu)}]\ln\left(\frac{\omega_J^2\tan^2(R/2)}{\mu^2}\right)-\gamma^{H}[\alpha_s{(\mu)}] \right) C_{i \to j}(z, \omega_J R, \mu) \,,
\end{align}
where $C_{i \to j}(z, \omega_J R, \mu)$ are the hard matching coefficients evolved from $\mu_J$ to $\mu_H$ via the time-like DGLAP equation. At NLO, these coefficients are given by
\begin{subequations}
	\label{eq:hardmatch}
	\begin{align}
		C_{q\to q'}(z, \omega_J R, \mu) & = \delta_{qq'}\delta(1-z) + \delta_{qq'}\frac{\alpha_s}{2\pi}\left[C_F\delta(1-z)\frac{\pi^2}{12} + P_{qq}(z)L \right. \notag \\[.2cm]
		                                & \left.- 2C_F(1+z^2)\left(\frac{\ln(1-z)}{1-z}\right)_+ - C_F(1-z)\right]\,,                                                   \\[.2cm]
		C_{q\to g}(z, \omega_J R, \mu)  & = \frac{\alpha_s}{2\pi}\left[(L-2\ln(1-z))P_{gq}(z) - C_Fz\right]\,,                                                          \\[.2cm]
		C_{g\to g}(z, \omega_J R, \mu)  & = \delta(1-z) + \frac{\alpha_s}{2\pi}\left[\delta(1-z)\frac{\pi^2}{12} + P_{gg}(z)L \right. \notag                            \\[.2cm]
		                                & \left.- \frac{4C_A(1-z+z^2)^2}{z}\left(\frac{\ln(1-z)}{1-z}\right)_+\right]\,,                                                \\[.2cm]
		C_{g\to q}(z, \omega_J R, \mu)  & = \frac{\alpha_s}{2\pi}\left[(L-2\ln(1-z))P_{qg}(z) - T_F2z(1-z)\right]\,,
	\end{align}
\end{subequations}
where $L=\ln\frac{\mu^2}{\omega_J^2\tan^2 R/2}$\,. With these components, the resummed EC jet function is obtained as
\begin{align}
	J_{c}(z_{c},\chi,\omega_{J}R,\mu_{J})
	 & =\sum_{h}\int dj_{\perp}2\pi j_{\perp}\int dz_{h}z_{h}\mathcal{G}_{c}^{h}(z_{c},z_{h},\omega_{J}R,j_{\perp},\mu_{J})\delta(\frac{2j_{\perp}}{z_{h}\omega_{J}}-\chi)
	\\
	 & =\frac{\pi\chi\omega_{J}^{2}}{2}C_{c \rightarrow i}(z_{c},\omega_{J}R,\mu_{J}) \int\frac{bdb}{2\pi}J_{0}\left(\frac{\chi\omega_{J}b}{2}\right)e^{-S_{pert}^{i}(\mu_b,\mu_{J})}
	\\
	 & \times\sum_j\int_{0}^{1}d\omega\omega \tilde{C}_{j\leftarrow i}(\omega,b,\mu_b,\nu_D,\nu_S)\sum_{h} \int_{0}^{1}d\hat{z_{h}}\hat{z_{h}} D_{h/j}(\hat{z_{h}},\mu_{b})\,.
\end{align}
where $S_{\mathrm{pert}}^i(\mu_b,\mu_J)$ is the perturbative sudakov factor
\begin{align}
	S_{\mathrm{pert}}^i(\mu_b,\mu_J)=\int_{\mu_{b}}^{\mu_J}\frac{d\mu^{\prime}}{\mu^{\prime}}\Bigg(\Gamma_{\mathrm{cusp}}^i\ln\bigg(\frac{\mu_J^2}{\mu^{\prime2}}\bigg)+\gamma^{D_i}+\gamma^{S_i}\Bigg)\bigg(\frac{\nu_D}{\nu_{S}}\bigg)^{\gamma^R[\alpha_s(\mu_b)]}\,,
\end{align}
The term $\tilde{C}_{j\leftarrow i}(\omega,b,\mu_b)$ represent matching from TMDFFs to collinear FFs, and $D_{h/j}(\hat{z_{h}},\mu_{b})$ are collinear fragmentation functions. For simplification in the ensuing discussion, we can represent the perturbative part of the TMDFFs matching coefficient as
\begin{align}
	\mathbf{D}_{i}^{\text{pert}}(b,\mu_b,\nu_D,\nu_S) =\sum_j\int_{0}^{1}d\omega\omega \tilde{C}_{j\leftarrow i}(\omega,b,\mu_b,\nu_D,\nu_S)\,,
\end{align}
where the one-loop expression at initial scale are~\cite{Luo:2019bmw,Luo:2019hmp}
\begin{subequations}
	\begin{align}
		\mathbf{D}_{q}^{\text{pert}}(b,\mu_b,\nu_D,\nu_S)
		 & = C_F \left( - 8\zeta_2 + 4 \right)\,,
		\\[.2cm]
		\mathbf{D}_{g}^{\text{pert}}(b,\mu_b,\nu_D,\nu_S)
		 & =  C_A \left( -8\zeta_2 + \frac{71}{18} \right) - n_f \frac{11}{18}\,.
	\end{align}
\end{subequations}
we note it as $\mathbf{D}_{i}^{\text{pert}}(b,\mu_b)$\,. Since $\ln R$ terms appear in $C_{i \to j}(z, \omega_J R, \mu)$, further resummation of these logarithms can be achieved by the time-like DGLAP evolution of the EC jet function form $\mu_J$ to $\mu$, governed by
\begin{align}
	\mu\frac{d}{d\mu}J_i(z,\chi,\omega_JR,\mu)=\frac{\alpha_s}{2\pi}\sum_j\int_z^1\frac{dz^{\prime}}{z^{\prime}}P_{ji}\left(\frac{z}{z^{\prime}}\right)J_j(z,\chi,\omega_JR,\mu)\,.
\end{align}
Next, to compute the full result at the desired precision, we consider the normalization details for the EC, The denominator $\sigma_0$ of the one inclusive cross-section form is expressed using the semi-inclusive Fragmenting Jet Functions (siFJFs) as~\cite{Kang:2016mcy}
\begin{align}
	\frac{d\sigma^{pp\to(\mathrm{jet})X}}{dp_T d\eta }  =
	 & \sum_{a,b,c}\frac{2}{p_T}\int_{z_0}^{1}\frac{dz_c}{z_c^2}\int_{VW/z_c}^{1-(1-V)/z_c}dv \int_{VW/vz_c}^1 d\omega x_a f_a (x_a,\mu) x_b f_b(x_b,\mu) \nonumber
	\\
	 & \times\frac{d\sigma^c_{ab}}{dvdw}(\hat{s},\hat{p}_T,\hat{\eta},\mu)\mathcal{J}_{c}(z_c,\omega_JR,\mu)\,,
\end{align}
where $\mathcal{J}(z,\omega_JR,\mu)$ is the renormalized siFJF, its NLO expressions are
\begin{subequations}
	\label{eq:sifjf}
	\begin{align}
		\mathcal{J}_{q}^{(1)}(z,\omega_{J} R,\mu)= & \frac{\alpha_{s}}{2\pi}L\left[P_{qq}(z)+P_{gq}(z)\right]-\frac{\alpha_{s}}{2\pi}\biggl\{C_{F}\left[2\left(1+z^{2}\right)\left(\frac{\ln\left(1-z\right)}{1-z}\right)_{+}+(1-z)\right] \nonumber \\
		                                           & -\delta(1-z)d_J^{q,\mathrm{alg}}+P_{gq}(z)2\ln{(1-z)}+C_F z\biggr\} \,,                                                                                                                         \\
		\mathcal{J}_g^{(1)}(z,\omega_J R,\mu)=     & \frac{\alpha_s}{2\pi}L\left[P_{gg}(z)+2n_f P_{qg}(z)\right]-\frac{\alpha_s}{2\pi}\biggl[\frac{4C_A(1-z+z^2)^2}{z}{\left(\frac{\ln(1-z)}{1-z}\right)}_{+} \nonumber                              \\
		                                           & -\delta(1-z)d_J^{g,\mathrm{alg}}+4n_f\left(P_{qg}(z)\ln(1-z)+T_F z(1-z)\right)\biggr]\,.
	\end{align}
\end{subequations}
Comparing the hard matching coefficients~\eqref{eq:hardmatch} with the siFJFs~\eqref{eq:sifjf}, the logarithms involving $R$ are expected to cancel in the normalized EC. This cancellation arises from the factorization property, which separates in-jet angular information from out-of-jet information. To illustrate, we rewrite the normalized EC schematically as
\begin{align}
	\mathrm{EC}(\chi)=\int\frac{\sigma\otimes J}{\sigma\otimes \mathcal{J}}dp_{T}d\eta\,,
	\label{eq:nor}
\end{align}
the normalization can be expanded perturbatively, up to NLO
\begin{align}
	\begin{aligned}
		\frac{\sigma \otimes J}{\sigma\otimes\mathcal{J}}=\frac{\sigma^{(0)}\otimes J^{(0)}+\sigma^{(0)}\otimes J^{(0)}+\sigma^{(0)}\otimes J^{(1)}+\sigma^{(1)}\otimes J^{(0)}}{\sigma^{(0)}\otimes\mathcal{J}^{(0)}+\sigma^{(0)}\otimes\mathcal{J}^{(1)}+\sigma^{(1)}\otimes\mathcal{J}^{(0)}}\,.
	\end{aligned}
\end{align}
Let $\sigma_0^{tot} = \sigma^{(0)}\otimes\mathcal{J}^{(0)} $ be the leading-order normalization factor. Using the relation
\begin{align}
	\sigma^{(1)}\otimes J^{(0)}=\frac{\left(\sigma^{(0)}\otimes J^{(0)}\right)\left(\sigma^{(1)}\otimes\mathcal{J}^{(0)}\right)}{\sigma_0^{tot}}\,,
\end{align}
the expression in Eq.~\eqref{eq:nor} will be
\begin{align}
	\mathrm{EC}(\chi)= & =\frac{1}{\sigma_0^{tot}}\sum_{a,b,c}\int  \frac{2}{p_T} d p_T\int d \eta \int_{z_{0}}^{1}\frac{1}{z_{c}^{2}} \int_{VW/z_c}^{1-(1-V)/z_c}dv \int_{VW/vz_c}^1 d\omega
	\nonumber\\
	                   & \times x_a f_a(x_a,\mu) x_b f_b(x_b,\mu) \frac{d\sigma^c_{ab}}{dvdw} J^{nor}_c(z_c,\chi,\omega_JR,\mu)\,.
\end{align}
Here, $J^{nor}$ differs from $J$ primarily in that its hard matching coefficients $C^{nor}_{c \rightarrow i}$ only contain the $\delta(1-z)d_J^{c,\mathrm{alg}}$ terms after the cancellation of the $L$ terms against similar terms from $\mathcal{J}$. Therefore, the DGLAP evolution dedicated to these $L$ terms have effectively canceled out in the calculation of the normalized EC. A further consequence is that, the contribution of the NLO partonic cross sections for the normalized EC will vanish. The resulting normalized EC is therefore expected to exhibit a significantly reduced dependence on the factorization scale $\mu$. Its primary remaining scale dependence will be associated with the jet scale $\mu_J$, characterized by $p_T R$.

\subsection{Non-Global Logarithms}
The EC Jet function, by its definition, focuses on measuring particle distributions and energy flow exclusively within the confines of a defined jet cone  of radius $R$. This inherent insensitivity to energy deposited far outside this cone categorizes the EC Jet function as a non-global observable~\cite{Dasgupta:2001sh,Dasgupta:2002bw,Banfi:2002hw}. Consequently, its precise theoretical description must account for NGLs. NGLs represent a significant class of large logarithmic corrections in QCD that arise when an observable is sensitive to emissions in such a restricted region of phase space. These logarithms typically originate from correlated soft wide-angle gluon emissions, where radiation initially directed outside the observed jet cone can subsequently radiate back into it, thereby influencing measurements.

To systematically address these logarithmic corrections, this work employs a formalism rooted in jet effective theory~\cite{Becher:2016omr,Becher:2016mmh,Becher:2015hka}, specifically drawing from developments for jet observables. This framework facilitates factorization with multi-Wilson line structure~\cite{Kang:2020yqw} and enables the simultaneous resummation of both standard "global" Sudakov logarithms and the more intricate NGLs.

Within the TMD region where $j_\perp \ll \omega_J R$, the factorized form of the siTMDFJF, generalized to account for NGLs, can be expressed as~\cite{Kang:2020yqw}
\begin{align}
	\mathcal{G}_c^h
	 & (z,z_h,\omega_JR,\boldsymbol{j}_\perp,\mu)=  \int d^{2}\boldsymbol{k}_{\perp}d^{2}\boldsymbol{\lambda}_{\perp}\delta^{2}\left(z_{h}\boldsymbol{\lambda}_{\perp}+\boldsymbol{k}_{\perp}-\boldsymbol{j}_{\perp}\right)
	\nonumber                                                                                                                                                                                                                                                       \\
	 & \times \sum_{m=2}^{\infty} \frac{1}{N_c} \mathrm{Tr}_c \left[ \mathcal{H}_m^i (\{\underline{n}\}, \omega_JR, \mu) \otimes \mathbf{S}_m (\{\underline{n}\}, \boldsymbol{\lambda}_{\perp}, \mu, \nu) \right] D_{h/i}(z_h, \boldsymbol{k}_{\perp}, \mu, \nu)\,.
\end{align}
Here, the multi-Wilson line structure that makes the hard and soft function matrices in the color space, $\mathrm{Tr}_c$ denotes a color trace. The symbol $\otimes$ indicates a convolution over the phase space of the parton directions $\{\underline{n}\}$~\cite{Becher:2016omr}
\begin{align}
	\mathcal{H}_m(\{\underline{n}\}) \otimes \mathbf{S}_m (\{\underline{n}\}, \boldsymbol{\lambda}_{\perp}) = \prod_{i=2}^{m} \int \frac{d\Omega(n_i)}{4\pi} \mathcal{H}_m(\{\underline{n}\}) \mathbf{S}_m(\{\underline{n}\})\,.
\end{align}
The squared matrix element for soft emissions of $m$ partons as
\begin{align}
	\mathbf{S}_m (\{\underline{n}\}, \boldsymbol{\lambda}_{\perp}) = \sum_{X_S} \langle 0 | \mathbf{S}_1^\dagger (n_1) \dots \mathbf{S}_m^\dagger (n_m) | X_J \rangle \langle X_J | \mathbf{S}_1 (n_1) \dots \mathbf{S}_m (n_m) | 0 \rangle \delta^2 (\mathcal{P}^{\epsilon J}_{\perp} - \boldsymbol{\lambda}_{\perp})\,.
\end{align}
The RG equations for the hard and in-jet soft functions, including NGL effects, are similar the RG equations of global logarithms, we can obtion the RG equations of hard functions and in-jet soft funcions
\begin{align}
	\frac{d}{d\ln\mu} \mathcal{H}_m(\{n\}, \omega_J, \mu) = \sum_{l=2}^{m} \mathcal{H}_l(\{n\}, \omega_J, \mu) \biggl\{\biggl[\Gamma_{\text{cusp}}(\alpha_s) & \ln\frac{\omega_J^2\tan^2 R/2}{\mu^2}
		- \gamma^{D}(\alpha_s)
	\nonumber                                                                                                                                                                                                                             \\ - \gamma^S(\alpha_s)\biggr]
	                                                                                                                                                         & \delta_{lm} \mathbf{1} - \hat{\mathbf{\Gamma}}_{lm}(\{n\}, \mu)\biggr\}\,,
\end{align}
\begin{align}
	\frac{d}{d\ln\mu} \mathbf{S}_l(\{n\}, b, \mu, \nu R) = \sum_{m=l}^{\infty} \biggl\{ \biggl[ -\Gamma_{\text{cusp}} \ln\frac{\nu^2 \tan^2 R/2}{\mu^2} + \gamma^S \biggr] & \delta_{lm} \mathbf{1} + \hat{\mathbf{\Gamma}}_{lm}(\{n\}, \mu) \biggr\}
	\nonumber                                                                                                                                                                                                                                         \\
	                                                                                                                                                                       & \hat{\otimes} \mathbf{S}_m(\{n\}, Q, \mu, \nu R)\,.
\end{align}
The $\hat{\otimes}$ means integrating over the $(m-l)$ additional directions of the out-jet partons, the structure of this result arise because the soft radiation is not constrained inside the jet.After solving the RG equations, we can obtain the resummation formula as
\begin{align}
	J_{c}(z_{c},\chi,\omega_{J}R,\mu_{J})
	=\frac{\pi\chi\omega_{J}^{2}}{2} & C_{c \rightarrow i}(z_{c},\omega_{J}R,\mu_{J})  \int\frac{bdb}{2\pi}J_{0}\left(\frac{\chi\omega_{J}b}{2}\right)e^{-S_{pert}^{i}(\mu_b,\mu_{J})}
	\nonumber                                                                                                                                                                          \\
	                                 & \times\mathbf{D}_{i}^{\text{pert}}(b,\mu_b) \hat{z_{h}} D_{h/j}(\hat{z_{h}},\mu_{b})U_{\text{NG}}(\mu_{b}, \mu_J)\,,
\end{align}
where the non-global evolution function $U_{\text{NG}}$, control the non-global logarithms, is given by
\begin{align}
	U_{\text{NG}}(\mu_{b}, \mu_J) = \frac{1}{N_c} \sum_{l=2}^{\infty} \text{Tr}_c \left[ \mathcal{H}_l (\{n'\}, \omega_J, \mu_J) \otimes \sum_{m \geq l}^{\infty} U_{lm} (\{n\}, \mu_J, \mu_b) \hat{\otimes} S_m (\{n\}, b, \mu_{b}) \right] \, ,
\end{align}
In many practical applications, $U_{\text{NG}}$ can be parameterized, in the framework of the lagre-$N_C$ and dipole angle integration, we can calculate the coefficients of the non-global single logarithms via  two-loop soft function as~\cite{Dasgupta:2012hg}
\begin{align}
	I & = 4 \int d\eta_1 \frac{d\phi_1}{2\pi} \int d\eta_2 \frac{d\phi_2}{2\pi}
	\nonumber                                                                                                                                                                                                                         \\
	  & \quad \times \frac{1 - e^{\eta_1} \cos \phi_1 - e^{\eta_2} \cos \phi_2 + e^{\eta_1+\eta_2} \cos(\phi_1 - \phi_2)}{(\cosh(\eta_1 - \eta_2) - \cos(\phi_1 - \phi_2)) (\cosh \eta_1 - \cos \phi_1) (\cosh \eta_2 - \cos \phi_2)}
	\nonumber                                                                                                                                                                                                                         \\
	  & \quad \times \Theta (R^2 - (\eta_2^2 + \phi_2^2)) \Theta (\eta_1^2 + \phi_1^2 - R^2)
	\nonumber                                                                                                                                                                                                                         \\
	  & = \frac{\pi^2}{3} + \alpha_2 R^2 + \alpha_4 R^4 + \mathcal{O}(R^6)\, ,
\end{align}
where $\alpha_2 \approx 0$ and $\alpha_4 \approx 0.013$, this result is exact for hemisphere jet masses in $e^{-}e^{+}$collisions, implies that the result obtained does not depend on the jet cone radius. Based on such calculations, a common parameterization, inspired by~\cite{Dasgupta:2001sh} can be used in the EC jet function, the parameterization is given by
\begin{align}
	U_{\text{NG}}(\mu_{b}, \mu_J) = \exp \left[ -C_A C_a \frac{\pi^2}{3} u^2 \frac{1 + (au)^2}{1 + (bu)^c} \right],
\end{align}
where the variable $u$ is given by
\begin{align}
	u = \int_{\mu_{b}}^{\mu_J} \frac{d\mu}{\mu} \frac{\alpha_s(\mu)}{2\pi} = \frac{1}{\beta_0} \ln \left[ \frac{\alpha_s(\mu_{b})}{\alpha_s(\mu_J)} \right],
\end{align}
and the constants are given as $a = 0.85 C_A$, $b = 0.86 C_A$, $c = 1.33$. This parameterization's precision is LL, the detailed work of NGL beyond the LL accuracy can be found in~\cite{Banfi:2021owj,Caron-Huot:2015bja}.

\subsection{Non-Perturbative And Hadronization Effects}
So far, our discussion has centered on the evolution of the EC Jet function primarily within the perturbative QCD regime. This approach is valid when the relevant momentum scales are large and the $b$ is small. However, the evolution of TMDs, particularly in the large-$b$ region, inherently incorporates significant non-perturbative contributions.

To address this, we adopt a common strategy for modeling the large-$b$ behavior, exemplified by the $b_\ast$-prescription~\cite{Collins:2014jpa,Collins:1984kg}. In this prescription, the impact parameter $b$ in perturbative calculations is effectively replaced by $b_\ast$, defined as
\begin{align}
	b_* = \frac{b}{\sqrt{1 + \frac{b^2}{b_{\text{max}}^2}}}\,.
\end{align}
Here, $b_{\text{max}}$ is a parameter that delineates the boundary between the perturbative and non-perturbative regions, we choose $b_\text{max} = 1.5$ as in~\cite{Kang:2015msa,Sun:2014dqm}. The $b_\ast$-prescription, and similar schemes, are designed to optimally integrate perturbative and non-perturbative aspects of TMD evolution. The primary goal is to maximize the applicability of perturbation theory in the small-$b$ region while systematically incorporating non-perturbative information where it becomes essential, typically at large $b$. This facilitates a smooth matching between the perturbative and non-perturbative descriptions within the TMD factorization formulation with $\mu_b \rightarrow \mu_{b_\ast}$.

Next, we consider the impact of hadronization corrections. TMDFFs are essential for describing the hadronization of a parton into an observed hadron. While the scale evolution of TMDFFs can be addressed using perturbative QCD, the actual process of hadron formation, the transition from a colored parton to a color-singlet hadron is intrinsically non-perturbative. To incorporate these crucial non-perturbative dynamics of hadronization into the TMDFF framework, dedicated model functions are introduced. The specific parameters of these hadronization models within TMDFFs are not derived from first principles but are instead determined by fitting comprehensive theoretical calculations (which combine perturbative evolution with these non-perturbative forms) to a wide array of experimental data, such as from SIDIS and Drell–Yan processes~\cite{Sun:2014dqm}.

To address the non-perturbative corrections within the TMD framework, we recall the perturbative Sudakov factor. Two primary components require non-perturbative modeling. First, the rapidity anomalous dimension necessitates a non-perturbative corrections, The non-perturbative corrections in soft rapidity anomalous dimension are implemented following the model~\cite{Li:2021txc} in
\begin{align}
	\frac{\nu_D}{\nu_S} \rightarrow \frac{\nu_D}{\nu_S}\times\text{exp}\bigg(-g_k(b)\ln\frac{\nu_D}{\nu_S}\bigg)\,.
	\label{eq:g}
\end{align}
Second, the TMD matching coefficients, which describe the transition from fragmentation functions to their TMD counterparts, also require non-perturbative input at large-$b$ region
\begin{align}
	S_i D_{i}(z, b; \mu_0, \nu_0) = S_i^{\text{pert}}  \mathbf{D}_{i}^{\text{pert}}(b,\mu_0,\nu_0)j_i(b)\,,
\end{align}
we can express the TMD model function as
\begin{align}
	j_i(b) = \left[ \mathbf{D}_{i}^{\text{pert}}(b,\mu_b) \right]^{-1}
	\sum_{h} \int_0^1 \, du \, dy \, (uy) \,
	\tilde{C}_{j\leftarrow i}(\omega,b,\mu_b)
	\, d_{h/i}(y, \mu_0) \, D_{h/i}^{\text{NP}}(uy, b)\,.
	\label{eq:j}
\end{align}
Combining these, we define a total non-perturbative function
\begin{align}
	S_i^{\text{NP}}(b) = j_i(b) \times \exp \bigg(-g_k(b)\ln\frac{\nu_D}{\nu_S}\bigg)\,.
\end{align}
For the normalized EC jet function, incorporating these non-perturbative effects, along with the $b_\ast$-prescription for the perturbative Sudakov factor and the non-global $U_{\text{NG}}$, leads to the complete normalized EC jet function, the resummed expression becomes
\begin{align}
	J^{\text{nor}}_{c}(z_{c},\chi,\omega_{J}R,\mu)
	=\frac{\pi\chi\omega_{J}^{2}}{2} & C^{\text{nor}}_{c \rightarrow i}(z_{c},\omega_{J}R,\mu) \int\frac{bdb}{2\pi}J_{0}\left(\frac{\chi\omega_{J}b}{2}\right)e^{-S_{pert}^{i}(\mu_{b_\ast}, \mu_{J})}
	\nonumber                                                                                                                                                                                          \\
	                                 & \times\mathbf{D}_{i}^{\text{pert}}(b,\mu_{b_\ast}) S_{i}^{\text{NP}}(b_\ast)U_{\text{NG}}(\mu_{b_\ast}, \mu_J)\,.
\end{align}

\section{Numerics and Comparison with Monte-Carlo Simulations}
\label{sec:results}
In this section, we present our numerical results of EC, and Monte Carlo simulation result using PYTHIA 8~\cite{Bierlich:2022pfr}, NLOJet$++$~\cite{Nagy:2005gn}(validate the $\alpha_s$ and $\alpha_s^2$ singularity in Fig.~\ref{fig:fixed} at $\chi \rightarrow 0$). We use the CT10nlo pdfset~\cite{Lai:2010vv} and the $\alpha_s$ in this set with LHAPDF 6~\cite{Buckley:2014ana}. we choose the anti-$k_T$ algorithm~\cite{Cacciari:2008gp} to reconstruct the jet. A common LO ansatz for $j_i(b)$~\eqref{eq:j} is
\begin{align}
	j_i(b) = \sum_h \int_0^1 dy \, y \, d_{h/i}(y, 1\,\text{GeV}) \, D^{\text{NP}}_{h/i}(y, b)\,,
\end{align}
where we use the DSS collinear fragmentation functions $d_{i/a}(y, 1\,\text{GeV})$~\cite{deFlorian:2007ekg,deFlorian:2007aj} and the TMDFF model funtion $D^{\text{NP}}_{h/i}(y, b)=\text{exp}(-0.042b^2/y^2)$. The TMD non-perturbative functions in~\eqref{eq:j} are
\begin{align}
	j_q(b) = \exp\big( -0.59b -0.03b^2 \big), \quad  j_g(b) = \exp\big( -0.17b -0.09b^2 \big),
\end{align}
and rapidity non-perturbative corrections in~\eqref{eq:g} is~\cite{Sun:2014dqm,Prokudin:2015ysa}
\begin{align}
	g_k(b) = - 0.21\frac{C_a}{C_F}\ln\frac{b}{b_\ast}\,.
\end{align}

\bef
\includegraphics[width=2.6in]{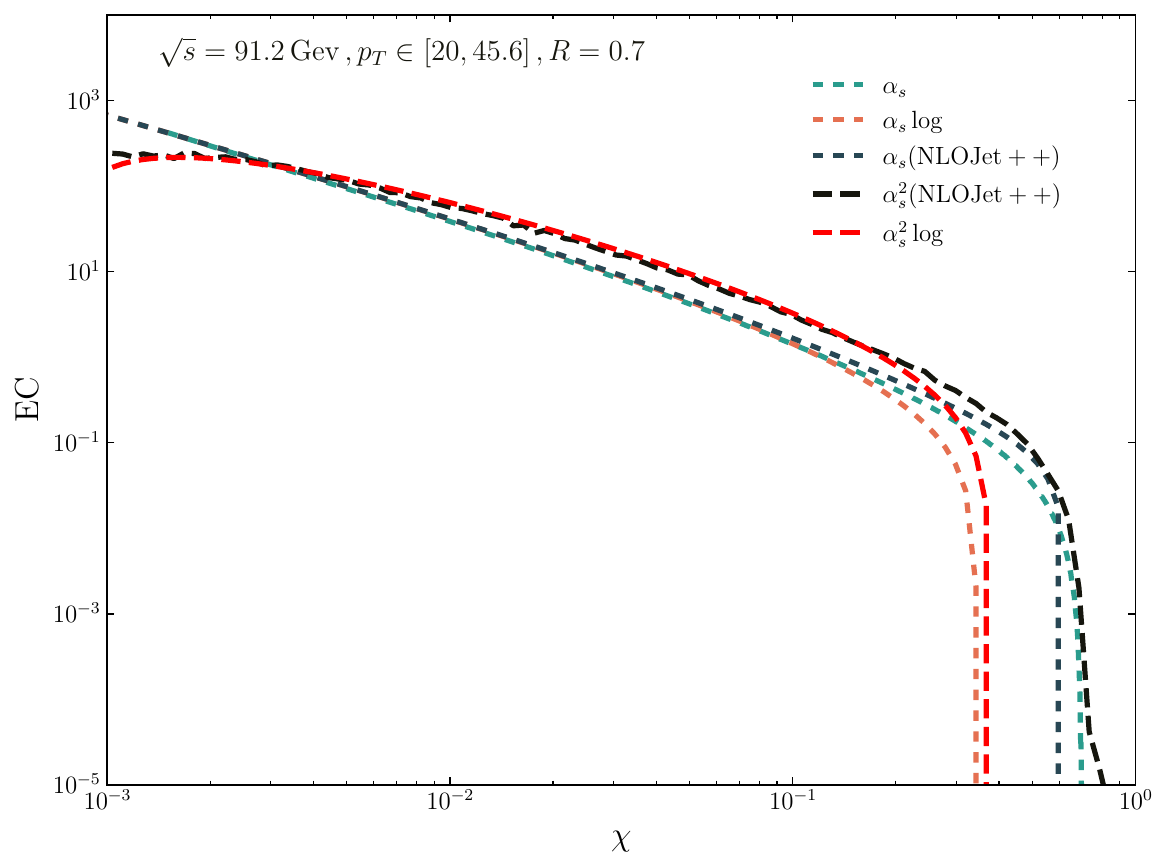}
\hskip 0.3in
\includegraphics[width=2.6in]{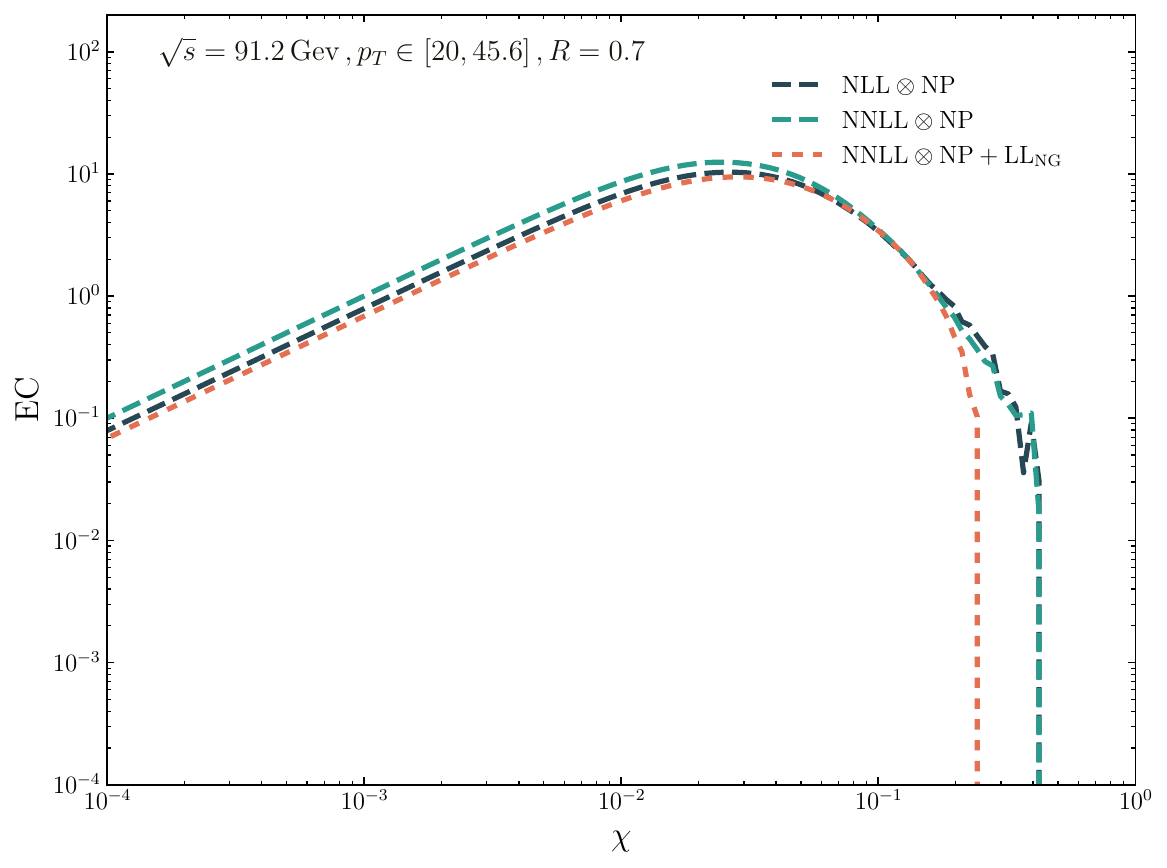}
\caption{Theoretical predictions for the normalized EC distribution at $e^{+}e^{-}$ annihilation. \textbf{Left Panel}: The comparison of the normalized EC distribution between the $\ln\chi$ singular contributions with the full fixed order calculations. Good agreement is observed at small angles. \textbf{Right Panel}: The effect of NGLs on the EC distribution, which is primarily confined to the small-angle region.}
\label{fig:fixed}
\eef

As a non-trivial validation of our factorization theorem of the normalized EC distribution, the left panel of Fig.~\ref{fig:fixed} compares three kinds of theoretical curves: (i) our prediction based on the factorization formula~\eqref{eq:fac}, which is dominated by leading-logarithmic $\ln\chi$ term; (ii) the full analytical fixed-order (FO) calculation from Appendix~\ref{app:B}; (iii) a numerical FO result generated with NLOJet$++$. The leading-logarithmic $\ln\chi$ in these calculation are in good agreement in the small-$\chi$ regime, and theoretical full fixed order results are valid in the region $\chi \leq R$ . The right panel of Fig.~\ref{fig:fixed} illustrates the impact of NGLs on the EC distribution, calculated at $\text{NNLL}$ accuracy which introduces a minor suppression in the small- $\chi$ region. However, this does not alter the overall scaling behavior of the EC distribution, which remains consistent with $\text{EC}(\chi) \sim \chi$, indicative of a relatively uniform distribution of energy relative to the jet axis.
\bef
\includegraphics[width=2.6in]{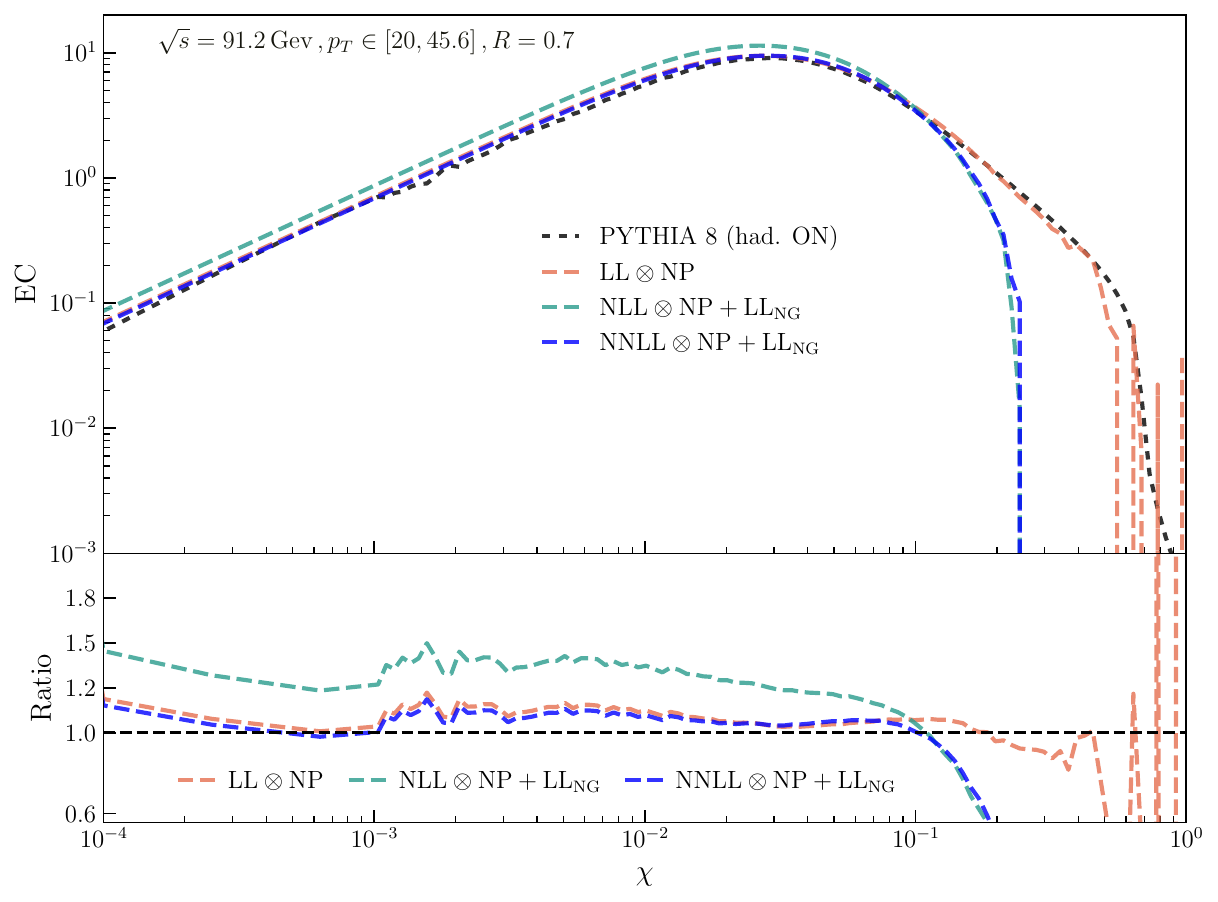}
\hskip 0.3in
\includegraphics[width=2.6in]{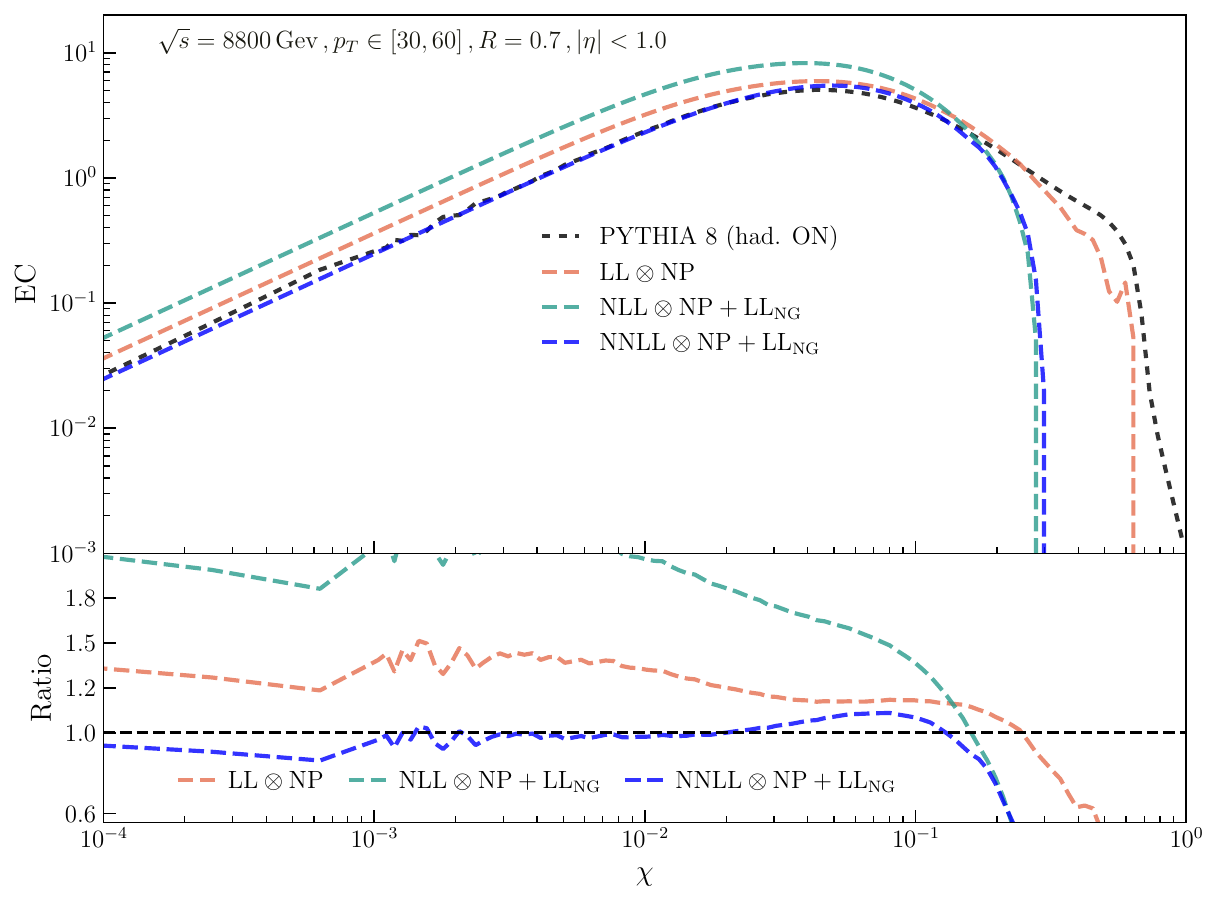}
\caption{The normalized EC distributions compared with PYTHIA simulations (hadronization enabled, only heavy hadrons decay) in both $e^{+}e^{-}$ annihilation and $pp$ collision at different theoretical accuracies. The results highlight EC's sensitivity and potential to constrain gluon TMD model parameters.}
\label{fig:ratio}
\eef

Fig.~\ref{fig:ratio} presents the normalized EC distribution at different theoretical accuracies, compared with PYTHIA simulations for both $e^{+}e^{-}$ annihilation and $pp$ collision. For the PYTHIA setup, hadronization and decay are turn on, and only heavy hadrons are allowed to decay. The left panel of Fig.~\ref{fig:ratio} focuses on $e^{+}e^{-}$ annihilation, where the EC is predominantly sensitive to quark TMDFFs. An good agreement is observed between our calculations (at both LL and NNLL+ $\text{LL}_{\text{NG}}$ accuracy) and the PYTHIA simulations. This concordance validates the description of quark TMD model within our framework. However, the EC in $pp$ collision is primarily governed by gluon TMDFFs, which are currently poorly constrained by experimental data~\cite{COMPASS:2013bfs,HERMES:2012uyd}. Compared to $e^{+}e^{-}$ annihilation, results from $pp$ collisions also exhibit greater sensitivity to model parameters. The right panel of Fig.~\ref{fig:ratio} focuses on $pp$ collisions, a great agreement at NNLL+NG accuracy. This observed sensitivity, particularly to the gluon TMD dynamics, suggests that the EC observable can be a valuable tool for constraining gluon TMD model parameters in future analyses.
\bef
\includegraphics[width=2.6in]{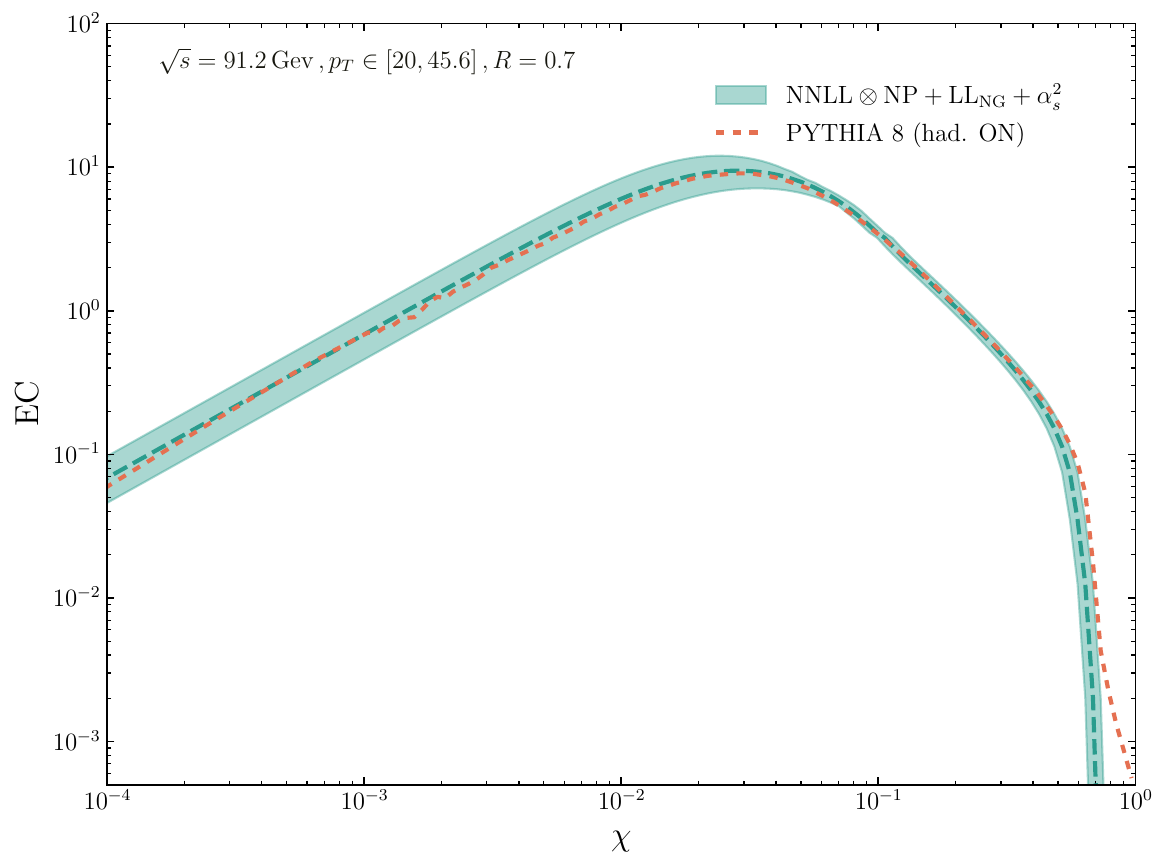}
\hskip 0.3in
\includegraphics[width=2.6in]{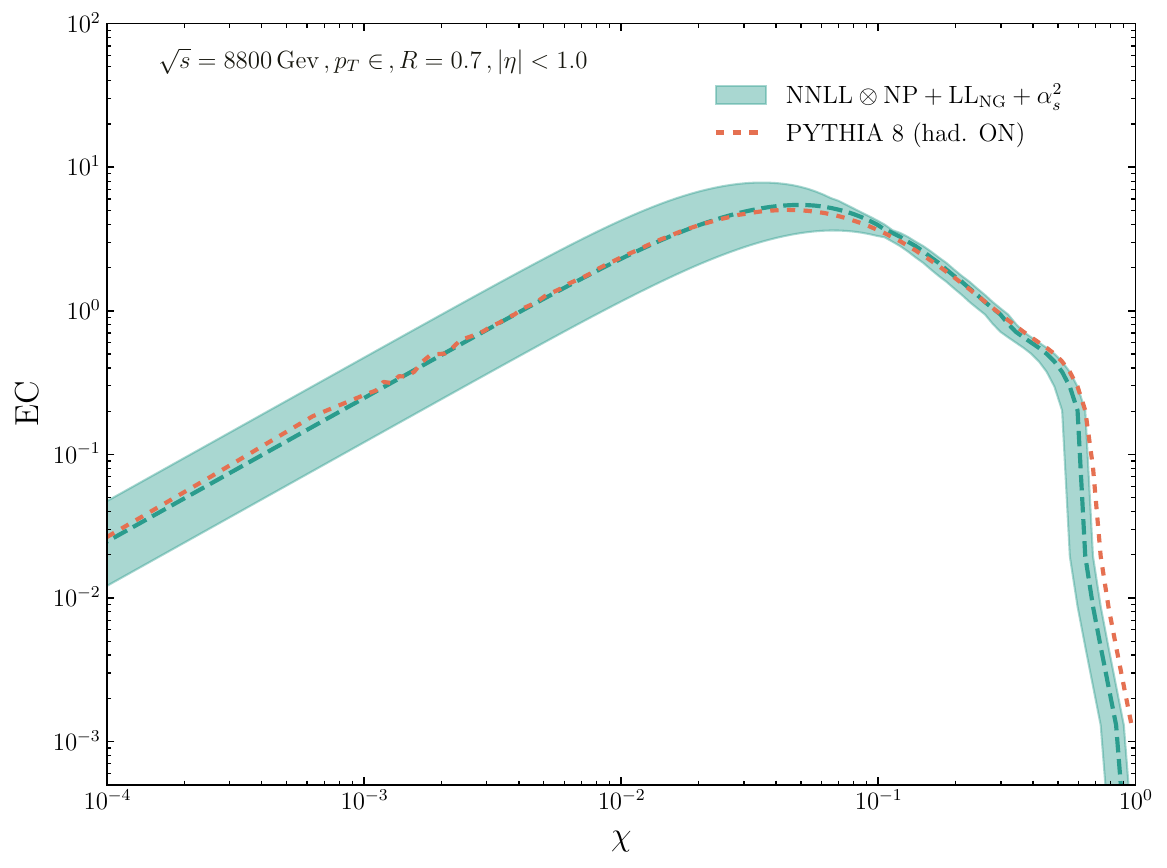}
\caption{Comparison of normalized EC distributions from theoretical predictions including uncertainty bands with PYTHIA 8 simulations, covering both $e^+e^-$ annihilation and $pp$ collisions.}
\label{fig:band}
\eef

In the region $\chi > 0.1$ , where the normalized EC distribution is dominated by the fixed-order result, we implement a matching scheme to ensure a smooth transition between the resummed and fixed-order calculations. The matched distribution is defined as
\begin{align}
    \text{EC}|_{\text{NNLL+NLO}} = t \bigg(\text{EC}|_{\text{NNLL}}+\big(\text{EC}|_{\text{NLO}} -\text{EC}|_{\text{exp. NLO}}\big)|_{\chi >0.1}\bigg)+(1-t)\text{EC}|_{\text{NLO}}
\end{align}
where $t(\chi)$ is the transition function
\begin{align}
    t(\chi) = \frac{1}{1+e^{a\frac{\chi-0.1}{0.1}}}
\end{align}
Fig.~\ref{fig:band}  illustrates our predictions for the normalized EC distribution based on this framework, including theoretical uncertainty bands, and compares them against PYTHIA 8 simulations for both $e^{+}e^{-}$ annihilation and $pp$ collision. The theoretical predictions are computed at $\text{NNLL}\otimes{\text{NP}}+\text{NLO}+\text{LL}_{\text{NG}}$ accuracy. The uncertainty bands are derived from variations of the resummation scale $\mu_J$ (from $\frac{1}{2} p_T R$ to $2p_T R$) and the factorization scale (from $\frac{1}{2} p_T$ to $2p_T$) . our result  smoothly matches the fixed-order result around $\chi \sim 0.1$, ensuring a correct description across a wide kinematic range. These combined comparisons provide a robust validation of our theoretical framework.

\section{Conclusion}
\label{sec:conclude}

In this work, we have introduced and developed a comprehensive theoretical framework for EC, a novel jet observable designed to precisely characterize the in-jet energy flow distribution. The primary motivation behind the EC is to furnish new insights into jet substructure and to pioneer a new avenue for the study of TMD physics, particularly for constraining gluon TMDFFs, which have proven notoriously difficult to extract experimentally.

Our theoretical formalism is established within SCET, leveraging the framework of siTMDFJFs. This robust approach meticulously incorporates TMD evolution, the resummation of large global logarithms, specifically $\ln(p_T R/\chi)$ up NNLL accuracy, and the resummation of NGLs at LL precision. Furthermore, crucial non-perturbative effects, including hadronization, are systematically addressed through established models and prescriptions like the $b_{\ast}$-formalism. The normalization procedure significantly mitigates the dependence of the EC on the factorization scale $\mu$, thereby rendering it a more refined observable that is predominantly sensitive to the intrinsic jet scale and effectively isolates the physics of the jet's internal structure.

We validate our theoretical framework by comparing our numerical predictions against Monte Carlo simulations from PYTHIA 8 for both $e^{+}e^{-}$ annihilation and $pp$ collisions. The comparison reveals good agreement in the small-$\chi$ region, the primary domain of validity for our factorization formalism. As expected, deviations emerge at large-$\chi$ region, consistent with the anticipated breakdown of the factorization approximation in this kinematic regime. To ensure a robust description across the full spectrum, this large-$\chi$ region is subsequently matched to the fixed-order calculation.

A significant outcome of this investigation is the clear demonstration of the EC's sensitivity to the underlying parton dynamics. In $e^{+}e^{-}$ annihilation, the EC predominantly probes quark TMDFFs, where our framework shows good concordance with simulations. More critically, in $pp$ collisions, the EC exhibits a pronounced sensitivity to gluon TMDFFs that are currently poorly constrained by global analyses of experimental data. This heightened sensitivity, particularly to gluon TMD dynamics, underscores the substantial potential of the EC. We therefore conclude that the one-point energy correlator is not only a valuable new tool for dissecting the intricate details of jet substructure but also represents a promising observable poised to provide significant new constraints on TMDFFs. Its application in future experimental analyses could be instrumental in advancing our understanding of the fundamental three-dimensional structure of hadrons and the complex QCD dynamics governing parton fragmentation with hadronization, especially for gluons within jets. Building on the promising sensitivity of the EC to TMDFFs, our ongoing work naturally leads us to consider how this observable might also illuminate the intricate world of TMD spin functions. While this study focused on unpolarized distributions, the EC's ability to precisely probe parton dynamics within jets suggests it could offer valuable insights into the spin-dependent aspects of hadronic structure.

\acknowledgments
We thank Xiaohui Liu for many inspiring and helpful discussions. We are also grateful to Feng Yuan and Hua Xing Zhu for valuable discussions and useful comments. This work is supported by the Natural Science Foundation China under Contract No. 12175016  and the Fundamental Research Funds for the Central Universities, Beijing Normal University.

\appendix
\section{Anomalous dimensions}
\label{app:A}
The NLO perturbative expressions for TMDFFs $D_{i/j}(z_h, \boldsymbol{b}, \mu, \nu)$ in $b$-space are~\cite{Kang:2017glf}
\begin{subequations}
	\begin{align}
		D_{q/q}(z_h, \boldsymbol{b}, \mu, \nu) & = \frac{1}{z_h^2} \Bigg\{ \delta(1-z_h) \notag                                                                                                                                                                                        \\
		                                       & \quad + \frac{\alpha_s}{2\pi}C_F\left[\frac{2}{\eta}\left(\frac{1}{\epsilon} + \ln\left(\frac{\mu^2}{\mu_b^2}\right)\right)+ \frac{1}{\epsilon}\left(\ln\left(\frac{\nu^2}{Q^2}\right) + \frac{3}{2}\right)\right]\delta(1-z_h) \notag \\
		                                       & \quad + \frac{\alpha_s}{2\pi}\left[-\frac{1}{\epsilon} - \ln\left(\frac{\mu^2}{z_h^2\mu_b^2}\right)\right]P_{qq}(z_h) \notag                                                                                                           \\
		                                       & \quad + \frac{\alpha_s}{2\pi}C_F\left[\ln\left(\frac{\mu^2}{\mu_b^2}\right)\left(\ln\left(\frac{\nu^2}{Q^2}\right)+ \frac{3}{2}\right)\delta(1-z_h) + (1-z_h)\right]\Bigg\}\,,
		\\[.2cm]
		D_{g/q}(z_h, \boldsymbol{b}, \mu, \nu) & = \frac{1}{z_h^2}\left\{\frac{\alpha_s}{2\pi}\left[-\frac{1}{\epsilon}-\ln\left(\frac{\mu^2}{z_h^2\mu_b^2}\right)\right]P_{gq}(z_h) + \frac{\alpha_s}{2\pi}C_Fz_h\right\}\,,
		\\[.2cm]
		D_{g/g}(z_h, \boldsymbol{b}, \mu, \nu) & =\frac{1}{z_h^2}\bigg\{\delta(1-z_h)   \notag                                                                                                                                                                                          \\
		                                       & \quad +\frac{\alpha_s}{2\pi}C_A\left[\frac2\eta\left(\frac1\epsilon+\ln\left(\frac{\mu^2}{\mu_b^2}\right)\right)+\frac1\epsilon\left(\ln\left(\frac{\nu^2}{Q^2}\right)+\frac{\beta_0}{2C_A}\right)\right]\delta(1-z_h) \notag          \\
		                                       & \quad +\frac{\alpha_s}{2\pi}\bigg[-\frac1\epsilon-\ln\bigg(\frac{\mu^2}{z_h^2\mu_b^2}\bigg)\bigg]P_{gg}(z_h)    \notag                                                                                                                 \\
		                                       & \quad +\frac{\alpha_s}{2\pi}C_A\bigg[\ln\bigg(\frac{\mu^2}{\mu_b^2}\bigg)\bigg(\ln\bigg(\frac{\nu^2}{Q^2}\bigg)+\frac{\beta_0}{2C_A}\bigg)\delta(1-z_h)\bigg]\bigg\}\,,
		\\[.2cm]
		D_{q/g}(z_h, \boldsymbol{b}, \mu, \nu) & =\frac{1}{z_{h}^{2}}\bigg\{\frac{\alpha_{s}}{2\pi}\bigg[-\frac{1}{\epsilon}-\operatorname{ln}\left(\frac{\mu^{2}}{z_{h}^{2}\mu_{b}^{2}}\right)\bigg]P_{qg}(z_{h})+\frac{\alpha_{s}}{2\pi}T_{F}2z_{h}(1-z_{h})\bigg\}\,.
	\end{align}
\end{subequations}
The NLO perturbative expressions for in-jet soft functions $S_i(\boldsymbol{b},\mu,\nu R)$ in $b$-space are~\cite{Kang:2017mda}
\begin{align}
	S_i(\boldsymbol{b},\mu,\nu R) & =1+\frac{\alpha_s}{2\pi}C_i\Bigg[\frac{2}{\eta}\bigg(-\frac{1}{\epsilon}-\ln\bigg(\frac{\mu^2}{\mu_b^2}\bigg)\bigg)+\frac{1}{\epsilon^2}-\frac{1}{\epsilon}\ln\left(\frac{\nu^2\tan^2(R/2)}{\mu^2}\right)
	\nonumber                                                                                                                                                                                                                                   \\
	                              & -\ln\bigg(\frac{\mu^2}{\mu_b^2}\bigg)\ln\bigg(\frac{\nu^2\tan^2(R/2)}{\mu_b^2}\bigg)+\frac12\ln^2\left(\frac{\mu^2}{\mu_b^2}\right)-\frac{\pi^2}{12}\Bigg]\,.
\end{align}
The cusp anomalous dimensions $\Gamma_{\mathrm{cusp}}$ and the non-cusp $\gamma$ have their usual expansion
\begin{subequations}
	\begin{align}
		\Gamma_{\mathrm{cusp}}\left[\alpha_s(\mu)\right]=\sum_{n=0}\Gamma_{\mathrm{cusp}}^n\left(\frac{\alpha_s(\mu)}{4\pi}\right)^{n+1},
		\\[2ex]
		\gamma_{D,S,R}\left[\alpha_s(\mu)\right]=\sum_{n=0}\gamma^{D,S,R}_{n}\left(\frac{\alpha_s(\mu)}{4\pi}\right)^{n+1},
	\end{align}
\end{subequations}
the QCD cusp anomalous dimension up to three loops are~\cite{Echevarria:2012pw}
\begin{subequations}
	\begin{align}
		\Gamma_{0}^{\mathrm{cusp}} = & 4C_{a},                                                                                       \\
		\Gamma_{1}^{\mathrm{cusp}} = & C_AC_a\bigg(\frac{268}{9}-8\zeta_2\bigg)-\frac{40C_an_f}{9},                                 \\
		\Gamma_{2}^{\mathrm{cusp}} = & C_A^2C_a\bigg(-\frac{1072\zeta_2}{9}+\frac{88\zeta_3}{3}+88\zeta_4+\frac{490}{3}\bigg)\notag \\
		                             & +C_AC_an_f\bigg(\frac{160\zeta_2}{9}-\frac{112\zeta_3}{3}-\frac{836}{27}\bigg)\notag         \\
		                             & +C_aC_Fn_f\bigg(32\zeta_3-\frac{110}{3}\bigg)-\frac{16C_an_f^2}{27},
	\end{align}
\end{subequations}
the soft anomalous dimension up to three loops are
\begin{subequations}
	\begin{align}
		\gamma_{0}^{S} = & 0                                                                                                                                                                 \\
		\gamma_{1}^{S} = & C_AC_a\left(\frac{22\zeta_2}{3}+28\zeta_3-\frac{808}{27}\right)+C_an_f\bigg(\frac{112}{27}-\frac{4\zeta_2}{3}\bigg)                                              \\
		\gamma_{2}^{S} = & C_{A}^{2}C_{a}\bigg(-\frac{176}{3}\zeta_{3}\zeta_{2}+\frac{12650\zeta_{2}}{81}+\frac{1316\zeta_{3}}{3}-176\zeta_{4}-192\zeta_{5}-\frac{136781}{729}\bigg)\notag  \\
		                 & +C_AC_an_f\bigg(-\frac{2828\zeta_2}{81}-\frac{728\zeta_3}{27}+48\zeta_4+\frac{11842}{729}\bigg)\notag                                                            \\
		                 & +C_aC_Fn_f\left(-4\zeta_2-\frac{304\zeta_3}{9}-16\zeta_4+\frac{1711}{27}\right)+C_an_f^2\bigg(\frac{40\zeta_2}{27}-\frac{112\zeta_3}{27}+\frac{2080}{729}\bigg),
	\end{align}
\end{subequations}
the rapidity anomalous dimension up to two loops are
\begin{subequations}
	\begin{align}
		\gamma_{0}^{R}  = & 0 ,                                                                         \\
		\gamma_{1}^{R}  = & C_{a}C_{A}\left(28\zeta_{3}-\frac{808}{27}\right)+\frac{112C_{a}n_{f}}{27},
	\end{align}
\end{subequations}
the collinear anomalous dimension up to two loops are~\cite{Li:2016ctv}
\begin{subequations}
	\begin{align}
		\gamma_{0}^{D_q}  =
		                    & -6C_F                                                                                                            \\[.2cm]
		\gamma_{1}^{D_q}  = & -2C_F\Bigg[C_F\bigg(\frac{3}{2}-2\pi^2+24\zeta_3\bigg)+C_A\bigg(\frac{17}{6}+\frac{22\pi^2}{9}-12\zeta_3\bigg) \\[.2cm]
		                    & +T_FN_f\bigg(-\frac{2}{3}-\frac{8\pi^2}{9}\bigg)\Bigg]                                                          \\[.2cm]
		\gamma_{0}^{D_g}  = & -2\beta_0                                                                                                        \\[.2cm]
		\gamma_{1}^{D_g}  = & C_{a}C_{A}\bigg(28\zeta_{3}-\frac{808}{27}\bigg)+\frac{112C_{a}n_{f}}{27}
	\end{align}
\end{subequations}
where $C_a=C_F$ for quark and $C_a=C_A$ for gluon.

\section{Fixed-Order}
\label{app:B}
In this case, the factorization of \eqref{eq:fac} will no work, a non-vanishing contribution arises only when both partons are inside the jet. Under this condition, the entire initial quark energy $\omega$, is transferred to the jet. Thus, if $\omega_J$ denotes the jet energy, then $\omega_J = \omega$, resulting in the momentum fraction $z = \omega_J / \omega = 1$. In contrast, the energy of the fragmenting parton, $\omega_h$, can be smaller than the jet energy $\omega_J$. Consequently, the ratio $z_h = \omega_h / \omega_J$ is generally less than $1$\,. For a splitting process $i \rightarrow jk$, where $j$ signifies the fragmenting parton, the one-loop bare siFJFin the MS scheme can be written as
\begin{align}
	G_{i, \text{bare}}^{jk,(1)}(z, z_h, \omega_J, \mu) = \delta(1-z) \frac{\alpha_s}{\pi} \frac{(e^{\gamma_E} \mu^2)^\epsilon}{\Gamma(1-\epsilon)} \hat{P}_{ji}(z_h, \epsilon) \int \frac{dq_\perp}{q_\perp^{1+2\epsilon}} \Theta_{\text{alg}}
\end{align}
where $\Theta_{\text{alg}} = \theta\big(z_h(1-z_h)\omega_J\tan\frac{R}{2}-q_\perp\big) $ make both partons are inside the jet for anti-kt algorithm. The formula for EC can now be derived. This is achieved by inserting a weighting factor corresponding to the fragmenting parton energy fraction and an angular delta function, and subsequently integrating over this energy fraction and summing over the fragmenting parton. The full fixed-order EC can thus be written as
\begin{align}
	J_{i, \text{bare}}(z,\omega_J, \mu) = \sum_j \delta(1-z) \frac{\alpha_s}{\pi} \frac{(e^{\gamma_E} \mu^2)^\epsilon}{\Gamma(1-\epsilon)} &\int_0^{1 - \frac{\tan\chi}{2 \tan \frac{R}{2}}} dz_h z_h \hat{P}_{ji}(z_h, \epsilon) \nonumber\\ &\times\int \frac{dq_\perp}{q_\perp^{1+2\epsilon}} \delta\bigg(\arctan\bigg(\frac{q_\perp}{ \frac{z_h \omega_j}{2}}\bigg) - \chi\bigg)
\end{align}
where $\Theta_{\text{alg}} = \theta\big(z_h(1-z_h)\omega_J\tan\frac{R}{2}-\frac{z_h\omega_J \tan\chi}{2}\big) = \theta\big( 1-\frac{\tan\chi}{2\tan\frac{R}{2}} -z_h \big) $\,, this result can be simplified by performing the integral over $q_\perp$ as
\begin{align}
	J_{i, \text{bare}}(z,\omega_J, \mu)
	 & = \sum_j \delta(1-z) \frac{\alpha_s}{2\pi^2} \frac{e^{\epsilon\gamma_E}}{\Gamma(1-\epsilon)} \frac{1}{\mu^2}\frac{\pi \omega_J^2(\tan\chi+\tan^3\chi)}{2}\nonumber \\
	 & \times\bigg(\frac{\mu^2}{(\frac{\omega_J \tan\chi}{2})^2}\bigg)^{1+\epsilon}
	\int_0^{1 - \frac{\tan\chi}{2 \tan \frac{R}{2}}} dz_h z_h^{1-2\epsilon} \hat{P}_{ji}(z_h, \epsilon)
\end{align}
For an initial quark, the final result is:
\begin{align}
	J_{q, \text{bare}}(z,\omega_J, \mu) & = \delta(1-z) \frac{\alpha_s}{2\pi^2}\frac{1}{\mu^2}\frac{\pi \omega_J^2(\tan\chi+\tan^3\chi)}{2}\nonumber \\&\times\bigg(\frac{\mu^2}{(\frac{\omega_J\tan\chi}{2})^2}\bigg)_+
	\bigg\{-\frac{3}{2}+\frac{3}{2}\frac{\tan\chi}{\tan\frac{R}{2}}
	-\frac{3}{8}\frac{\tan^2\chi}{\tan^2\frac{R}{2}} -\log\bigg(\frac{\tan^2\chi}{4\tan^2\frac{R}{2}} \bigg)  \bigg\}.
\end{align}
Focusing on the singular logarithmic term above, in TMD region, this term can be decomposed into two parts
\begin{align}
	\chi \rightarrow 0 \quad \quad \log\bigg(\frac{\tan^2\chi}{4\tan^2\frac{R}{2}} \bigg) \approx -\log\bigg( \frac{\mu^2}{(\frac{\omega_J \chi}{2})^2}\bigg) + L\,,
\end{align}
the terms match the corresponding logarithmic structure found in the factorization formula for in-jet function and collinear function as shown in Fig.~\ref{fig:fixed}. The singularity at $\chi \rightarrow 0$ is the same as the factorization formula, which is consistent with the expectation that the EC distribution should reduce to the in-jet soft function and collinear function in this limit. The non-singular terms are not captured by the factorization formula, and they are not relevant for the TMD region.

\bibliographystyle{JHEP}
\bibliography{ExportedItems}
\end{document}